**Discrete Hall resistivity contribution from Néel skyrmions in multilayer nanodiscs**

Katharina Zeissler[1*], Simone Finizio[2], Kowsar Shahbazi[1], Jamie Massey[1], Fatma Al Ma'Mari[1,3], David M. Bracher[2], Armin Kleibert[2], Mark C. Rosamond[4], Edmund H. Linfield[4], Thomas A. Moore[1], Jörg Raabe[2], Gavin Burnell[1], and Christopher H. Marrows[1]

[1]*School of Physics and Astronomy, University of Leeds, Leeds LS2 9JT, United Kingdom*

[2]*Swiss Light Source, Paul Scherrer Institut, CH-5232 Villigen PSI, Switzerland*

[3]*Department of Physics, Sultan Qaboos University, 123 Muscat, Oman*

[4]*School of Electronic and Electrical Engineering, University of Leeds, Leeds LS2 9JT, United Kingdom*

*Correspondence to k.zeissler@leeds.ac.uk

Magnetic skyrmions are knot-like quasiparticles. They are candidates for non-volatile data storage in which information is moved between fixed read and write terminals. Read-out operation of skyrmion-based spintronic devices will rely upon electrical detection of a single magnetic skyrmion within a nanostructure. Here, we present Pt/Co/Ir nanodiscs which support skyrmions at room temperature. We measured the Hall resistivity whilst simultaneously imaging the spin texture using magnetic scanning transmission x-ray microscopy (STXM). The Hall resistivity is correlated to both the presence and size of the skyrmion. The size-dependent part matches the expected anomalous Hall signal when averaging the magnetisation over the entire disc. We observed a resistivity contribution which only depends on the number and sign of skyrmion-like objects present in the disc. Each skyrmion gives rise to 22±2 nΩ cm irrespective of its size. This contribution needs to be considered in all-electrical detection schemes applied to skyrmion-based devices.

The non-trivial topology of a skyrmion leads to an improved stability against external perturbations[1, 2, 3, 4, 5]. The topology of the spin texture is characterised by the skyrmion winding number $S$, which takes integer values[6]. Mathematically, $S$ is given by [1],

$$S = \frac{1}{4\pi} \int \boldsymbol{m} \cdot \left( \frac{\partial \boldsymbol{m}}{\partial x} \times \frac{\partial \boldsymbol{m}}{\partial y} \right) dx dy, \qquad (1)$$

where **m** is a unit vector pointing along the local magnetisation direction. Individual cylindrical skyrmions have a winding number of ±1. $S$ is an integer invariant for all mutually continuously deformable skyrmions. Hence, the associated topological Hall effect is also invariant under continuous deformation. A consequence of this, in the context of Néel skyrmions, is that the winding number is magnetic domain size and shape independent as long as the domain is wrapped by an unbroken Néel domain wall [1]. In multilayer films the formation of Néel type skyrmions is energetically favourable due to perpendicular magnetic anisotropy, interface Dzyaloshinskii-Moriya and dipolar interaction [7, 8, 9, 10, 11]. In thin film multilayers they are stable at room temperature [11, 12, 13] and can be moved using spin-torques [14, 15, 16, 17, 18], with the expectation that the current densities needed can eventually be as low as the $10^6$-$10^7$ A/m$^2$ observed in cryogenic measurements on single crystal materials [19, 20, 21]. This permits very low energy manipulation of information[22]. Whilst current-induced movement of skyrmions is needed to manipulate the encoded information, electrical detection of single skyrmions is essential to read back the stored information.

Here we have measured the Hall resistance of Pt/Co/Ir multilayer discs of 1 µm diameter while simultaneously imaging the magnetic configuration within it using STXM with x-ray magnetic circular dichroism (XMCD) contrast. STXM is a high resolution imaging technique which is magnetically non-invasive. The conclusions that we draw are twofold. The variation in skyrmion size with field leads to a signal that matches that expected from the anomalous Hall effect tracking the change in magnetisation. Furthermore, an additional magnetisation-independent Hall signal of 22±2 nΩ cm associated with the presence, number, and sign of individual skyrmion-like objects was found. Its size is comparable to the anomalous Hall signal for skyrmions with a radius smaller than 150±30 nm. This skyrmion number-dependent resistivity cannot qualitatively be explained with conventional Berry phase theories of the topological Hall effect, suggesting that further theoretical development is needed.

Skyrmions and skyrmion-like domains were stabilised in an electrically connected 1 μm diameter discs from [Pt/Co/Ir]$_{x10}$ multilayer stacks (see supplementary information, Fig. S1, for device, magnetic and electrical information). A current and field nucleation protocol was used to nucleate an integer number of skyrmions in the disc (see methods section and supplementary information, Fig. S2 for details). The Hall resistance was measured as a function of out-of-plane field. From the XMCD contrast images the normalised magnetisation, $M_z/M_{Sat}$, was computed by counting the number of black, white, and grey pixels in the disc and assigning them a value of 1, -1, and 0 respectively and normalising the sum by the number of pixels in the disc (see supplementary information for details, Fig. S3). Fig. 1a - c shows the electrical and magnetic response to an external out of plane field of one skyrmion, $N$ = 1, stabilised in the discs. The number $N$ is a counter of enclosed magnetic domains without drawing conclusions on their topology. In case of a skyrmion, $N$ = -$S$ is the sign convention adopted throughout this paper. The distinction between $S$ and $N$ becomes relevant when considering magnetic domains which touch the edge in such a way that the Néel domain wall wrapping is incomplete or when the wrapping is disrupted locally by, for example, a Bloch point. Measuring the Hall resistance throughout magnetic minor loops enabled the observation of changes in the resistance reflected by changes in the magnetic state while keeping the number of skyrmions in the disc constant was kept constant. The Hall resistance, the corresponding skyrmion diameter and normalised disc magnetisation throughout two minor loops are shown in Fig. 1a - c alongside XMCD images of the domain states d - g and h - i. In the final field sweep of Repeat 1 the applied field was increased until the skyrmion was annihilated (Major loop 6, Fig. 1a - c), which took place between 65 and 70 mT and reduced the value of $N$ from 1 to 0. After the skyrmion is annihilated the Hall resistivity returns to its saturation value of around -223±6 nΩ cm. A clear resistivity change with respect to the saturation value was observed for skyrmions with a diameter above 75 nm. XMCD images of skyrmion states corresponding to $N$ = -2, 2 and 3 are seen in Fig. 2 a - c, e - g and i - k respectively with the normalised magnetisation $M_z/M_{Sat}$ and the change in the normalised Hall resistance $R_{xy}/R_{xy,Sat}$ response d, h and l respectively. $R_{xy}$ refers to the measured Hall resistance and $R_{xy,Sat}$ is the Hall resistance at the saturated state corresponding to a disc with a completely white XMCD contrast. See supplementary information movies for all measured XMCD images.

In magnetic materials with topologically non-trivial spin textures, the Hall resistivity is given by the sum of the ordinary ($\rho_{xy}^N$), the anomalous ($\rho_{xy}^A$), and the topological ($\rho_{xy}^T$) Hall resistivities [6],

$$\rho_{xy} = \rho_{xy}^N + \rho_{xy}^A + \rho_{xy}^T = R_0 B + R_S \mu_0 M_z + R_0 P B_{eff}^z, \qquad (2)$$

where $R_0$ is the ordinary Hall coefficient, $B$ is the external applied magnetic field, $R_S$ is the anomalous Hall coefficient, $M_z$ is the z-component of the magnetisation, $P$ is the spin polarisation of the conduction electrons, and $B^z_{eff}$ is the effective field experienced by the conduction electrons due to Berry phase they accumulate whilst adiabatically traversing skyrmions [23, 24]. This magnetic field emerges as there is one quantum of emergent magnetic flux associated with each skyrmion [6, 25]. In this form, Equation 2 neglects any other contributions such as the planar Hall effect, which can arise due to anisotropic magnetoresistance mechanism.

Fig. 3a shows the nomalised Hall resistance $R_H/R_{H,Sat}$ against the normalised magnetisation $M_z/M_{Sat}$. The ordinary Hall resistance contribution was calculated using the measured ordinary Hall coefficient of -1.9±0.2×$10^{-11}$ Ωm/T and was subtracted from the measured value $R_{xy}$ to yield $R_H$. Therefore, in accordance with equation 1, changes in $R_H$ arise only from changes in the magnetisation and/or spin texture.

Eq. 2 shows that the anomalous Hall resistivity $\rho_{xy}^A$ is expected to be proportional to the *z*-component of the magnetisation, the same component that gives rise to XMCD contrast in the STXM setup. This condition is true if there are no changes in the longitudinal resistivity, since $R_s$ typically has a dependence on $\rho_{xx}$. As different magnetic states are measured in the same discs and at the same temperature, and the magnetoresistance is vanishingly small, this condition holds. The line connecting the two saturated states was fitted, shown in Fig. 3a yielding a slope of -1.01±0.01 and an intercept of 0.01±0.05. This slope is within the error bar of -1; the value expected on the basis of the Hall signal being purely due to an anomalous Hall effect arising from the magnetisation averaged over the area of the disc. All other data for $N > 0$ and $N < 0$ run parallel to this line i.e. retaining a slope of -1 (see Fig. 3b for the fitted slopes and intercepts). All data sets are within one error bar of a -1 slope except the $N = -2$ data set which is -1 within two error bars. This distribution is consistent with Gaussian statistics. We can therefore identify this variation in the Hall signal as the anomalous Hall effect associated with the skyrmion being compressed and expanded by the field. However, each data set is offset from the $N = 0$ line i.e. each $N \neq 0$ data set has a non-zero intercept indicating that there must be an additional resistivity contribution which is magnetisation independent.

The intercepts were converted into resistivities, $\rho_{xy}^{Int}$, and are plotted versus $N$ in Fig. 3b. $\rho_{xy}^{Int}$ is proportional to the number of skyrmion-like objects $N$ in the disc. The dashed line shows the linear fit through the data points and reveals that each skyrmionic object contributes 22±2 nΩcm to the measured resistivity.

Fig. 3c shows $\rho_{xy}^{Int}$, with respect to sequential image numbers. This highlights the skyrmion number dependent incremental change away from the $N = 0$ case. The XMCD images shown in Fig. 3 d to o highlight the magnetisation independence of the measured offset. The offset is the same for the same number of skyrmions $N$ even when they have very different sizes, compare the states depicted in Fig. 3d and e, for instance. A sharp drop in the measured offset is observed when one of the following two magnetic field driven scenarios occur (i) the individual skyrmions expand and fuse together (see Fig. 3c data points enclosed in the blue box and its corresponding magnetic state Fig. 3f) and (ii) the skyrmions are annihilated (see Fig. 3c data points enclosed in the green and red box, the corresponding magnetic state of the latter can be seen in Fig. 3i).

Images within the $N = 3$ data set indicate that there is a possibility that the magnetic domains are touching the edge for certain fields (see Fig. 2i – k). The imaging resolution and the parasitic x-ray absorption at the electrodes makes it hard to identify whether the domain is continuously wrapped by a Néel domain wall or whether the domain wall has been ejected at the edge, i.e. whether $S = N$ or whether $S \neq N$. Whilst the STXM shows only the out-of-plane component of the magnetisation, imaging of the in-plane magnetisation and hence the confirmation of Néel type domain walls in these multilayer systems was done using x-ray photoemission electron microscopy on films grown at the same time as the disc samples. Left-handed Néel domain walls were observed (see supplementary information Fig. S4). This is in agreement with Lorentz microscopy under tilted imaging conditions on closely related material grown in the same sputter system[26], as well as previous results from Chauleau et al. on multilayers with the same stacking sequence [27]. Such domain walls form topologically non-trivial skyrmionic spin textures when appearing as closed loops. No change in $\rho_{xy}^{Int}$ was observed until the magnetic domains were seen to annihilate or coalesce (see Fig. 3c green, red and blue boxes and f) from which we conclude that either the domains remain completely wrapped by Néel domain walls and that $S = N$, or that the physical origin of the magnetisation independent contribution is not reliant on an integer value for $S$.

Returning again to Eq. 2, it is tempting to identify this signal that occurs when the topology changes as the topological Hall effect $\rho_{xy}^T$, since the emergent field is proportional to the areal winding number density in that theory. The usual theory of that effect states that the change in the resistivity associated with the presence of skyrmions can be estimated using

$$\rho_{xy}^T \approx PR_0 \left|\frac{q^e}{e}\right| B_z^e = PR_0 \phi_0 S/A, \tag{3}$$

where $q^e$ is the emergent topological charge equal to ± ½ for different spins, and $B_z^e$ is the emergent magnetic field given by $(\hbar/2)8\pi\phi_z$ ($\phi_z$ is the skyrmion density $S/A$), $A$ is the area of the disc, and $\phi_0$ is the flux quantum $h/e$ with $h$ being Planck's constant and $e$ the electron charge [20, 23, 28]. In our case, using a maximum possible spin polarization of 100 %, $\rho_{xy}^T$ can be computed from this expression to have an upper limit of 0.01 nΩ cm per skyrmion in the disc. This value, despite representing the maximum expected by current theory, is far too small to account for the observed signal. On the other hand, other effects might be involved. Distortions in the current flow around the skyrmion due to the current-focussing effects arising from the anisotropic magnetoresistance (AMR) [29], the planar Hall effect [30], or d.c. eddy current loops [31] may also play a role in generating transverse voltages. Contributions from the planar Hall effect can be ruled out as it is a consequence of AMR. Local changes of the in plane magnetisation with respect to the current direction result in a change of the resistivity which then leads to a redistribution of the current density and hence the voltage equipotential measured. The AMR is even in magnetisation and hence $N$ = 2 and $N$ = -2 states (such as those shown in Fig. 3h and n) would show the same sign of planar Hall effect contribution, whilst an opposite sign is observed. Furthermore, the measured AMR ratio for the multilayer system is 0.04 %, whilst an effect at the level of 15% is required to generate a planar Hall resistivity at the level of the observed $\rho^{Int}_{xy}$ (see supplementary information for details, Fig. S5). D.c. eddy current loops are again a consequence of AMR and hence cannot produce the required magnitude. Eddy current loops are present in 180° walls, however a skyrmion can be viewed as a 360° wall i.e. two 180° walls, one up-down and one down-up, back-to-back, and hence any d.c. eddy current loops will cancel. Furthermore, any distortions in the current distribution caused by the AMR will be skyrmion-size dependent, whilst we see a constant $\rho^{Int}_{xy}$ for rather different skyrmion sizes with the same $N$ (see for instance states in Fig. 3d and e). Another explanation is therefore needed, and our results challenge conventional models of the Hall effects generated by skyrmions.

The Berry phase theories above treat systems that are chemically homogeneous and have an almost uniform skyrmion winding number density arising from a skyrmion lattice. Neither of these conditions are true in our multilayer system containing discrete skyrmions in a uniform background. Moreover the transport is unlikely to be adiabatic [32, 33], and the skyrmion will not be considerably smaller than the spin diffusion length in the metals from which our multilayer is formed [34, 35], as is commonly assumed in such theories. Hence, a new treatment of these real space Berry phase effects in such multilayer systems is needed to clarify these points.

## CONCLUSION

Electrical transport measurements have frequently been employed to study skyrmions in bulk-DMI non-centrosymmetric transition metal compounds while the material was in the skyrmion crystal phase [6]. Measurements were performed on bulk samples [36], single crystals [20, 23, 25], thin films [37, 38], and nanostructures [39, 40, 41]. Spin polarised scanning tunneling microscopy and differential tunnelling non-collinear magnetoresistance at low temperature has been shown to allow the detection of single skyrmions in ultra-high vacuum conditions [42, 43]. Readout with conventional magnetic tunnel junctions has been analysed theoretically but not practically demonstrated [44]. As for room temperature detection, an anomalous Hall detection scheme was showcased at room temperature which confirmed that the anomalous Hall resistance can detect a single skyrmion [45]. By counting the skyrmion number and comparing it to the Hall resistivity conflicting conclusions have been drawn.

On one side, only the ordinary and anomalous Hall contribution could be observed [45], while on the other side of the argument an additional, possibly topological, contribution was observed [46].

We have performed a combined imaging and electrical transport study that allows the direct correlation of the Hall signal in a skyrmion-bearing nanostructure with the exact spin texture that gives rise to it. We have used this method to separately determine the Hall signals associated with the presence and size of skyrmions. Whilst the latter dependence arises from the anomalous Hall Effect, the former requires a new explanation. It may have the same origins as that recently reported by Raju et al. [46] Nevertheless, the magnetisation independence and the observed proportionality to the number of skyrmions in the disc are a good indication that the origin is linked to the non-trivial topology of the observed magnetic structures.

The size independent contribution is of great technological importance as the expected anomalous Hall resistivity associated with a skyrmion drops below the $\rho^{Int}_{xy}$ = 22±2 nΩ cm contribution when the radius is reduced below 150±30 nm (see Fig. 3b). Understanding the origin of this contribution should lead to schemes by which it may be increased to still larger amplitudes. The ability to electrically detect single skyrmions not only opens up ways to study their static and dynamic properties in nanostructures under a wide range of experimental conditions, but also offers the prospect of a simple electrical method of reading out the state of future skyrmion-based spintronic devices.

**ACKNOWLEDGEMENTS**

Support from European Union (H2020 grant MAGicSky No. FET-Open-665095.103 and FP7 ITN "WALL" (Grant No. 608031)), as well as from the Diamond Light Source and the Swiss Nanoscience Institute (grant P1502), is gratefully acknowledged. Part of this work was carried out at the PolLux (X07DA) and SIM (X11MA) beamline of the Swiss Light Source. The PolLux end station was financed by the German Minister für Bildung und Forschung (BMBF) through contracts 05KS4WE1/6 and 05KS7WE1.

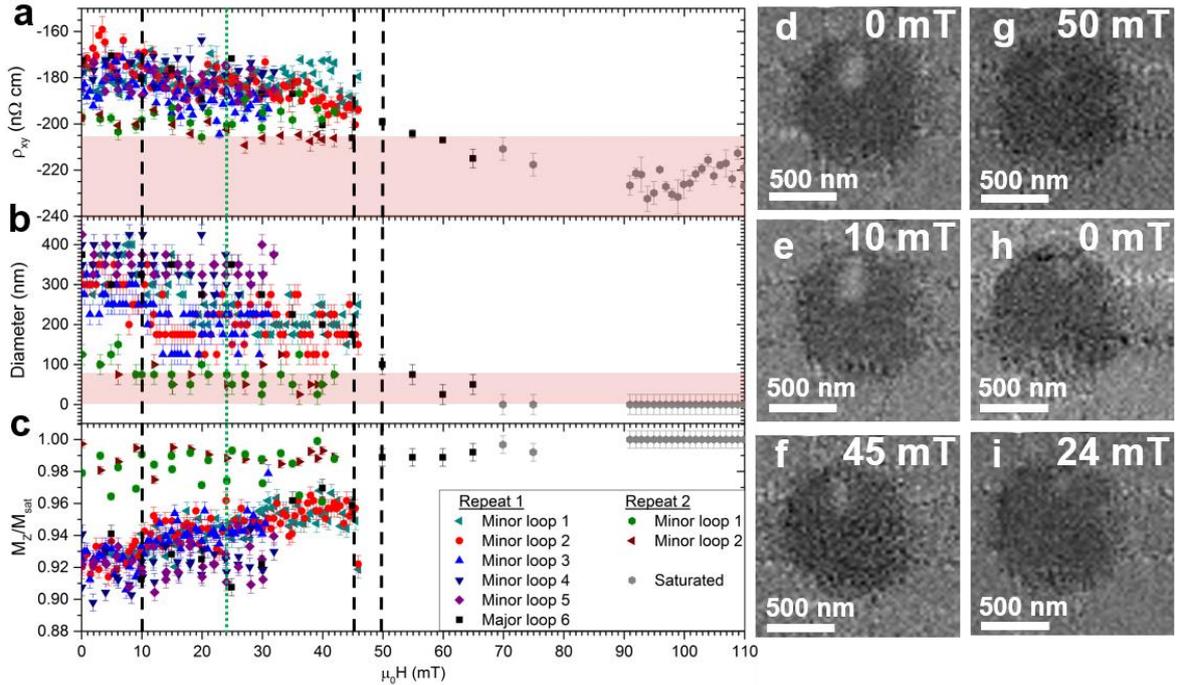

**Figure 1: Field evolution of the magnetic skyrmion.** The skyrmion measured was created using the current pulse/field protocol shown in Fig. S2. The skyrmion response to field is shown via the changes in **a**, Hall resistivity, $\rho_{xy}$, **b**, skyrmion diameter (error determined by the image resolution), and **c**, the out of plane magnetisation changes $M_z/M_{Sat}$ (the error is given by the counting statistics see supplementary information). The minor loop was repeated six times, with sixth cycle extending to high enough field, 75 mT, to finally annihilate the skyrmion. The red shaded area in **a** and **b** highlights skyrmions with a diameter of less than 75 nm which cannot be distinguished unambiguously from the saturated state using resistivity measurements. **d**, Shows a XMCD image of the skyrmion created and shown in Fig. S2 at 0 mT. **e-g** XMCD images of the skyrmion at 10 mT, 45 mT and 50 mT respectively, black dashed lines in **a** to **c**. The resistivity, diameter and magnetisation states correspond to the black squares and are highlighted by the black dashed lines. After the annihilation of the skyrmion shown in **d-g**, the system was returned to the saturated state and a new skyrmion was nucleated using the same current pulse and field creation protocol. **h** and **i** The XMCD image of the skyrmion at 0 mT and 24 mT, green dotted line in **a** to **c**. The resistivity, diameter and magnetisation state correspond to the green circles and the 24 mT state is highlighted by the green dotted line. Skyrmions larger than to 75 nm can be detected electrically.

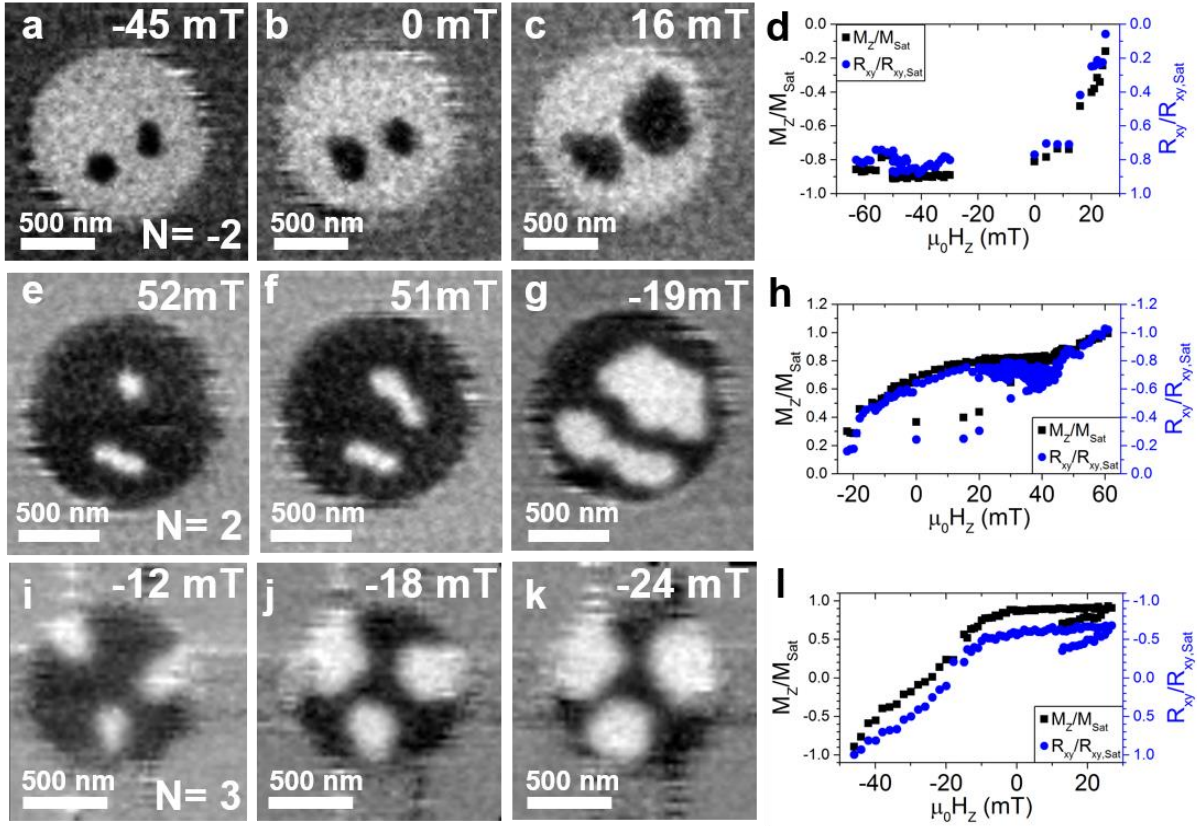

**Figure 2: Multiple magnetic skyrmion creation.** Using the current and field protocol two skyrmions and three skyrmions were created. **a-c** XMCD images of two skyrmions, *N* = -2, at different stages of the minor field loop. **d** Shows the measured normalised resistance, $R_{xy}/R_{xy,Sat}$, and the extracted normalised magnetisation $M_Z/M_{Sat}$. **e-g** Show XMCD images of two skyrmions embedded in an oppositely magnetised disc, *N* = 2, at different stages of the minor field loop. **h** Shows the measured normalised resistance, $R_{xy}/R_{xy,Sat}$, and the extracted normalised magnetisation $M_Z/M_{Sat}$. **i-k** Show XMCD images of three skyrmions, *N* = 3, at different stages of the minor field loop. **l** Shows the measured normalised resistance, $R_{xy}/R_{xy,Sat}$, and the extracted normalised magnetisation $M_Z/M_{Sat}$.

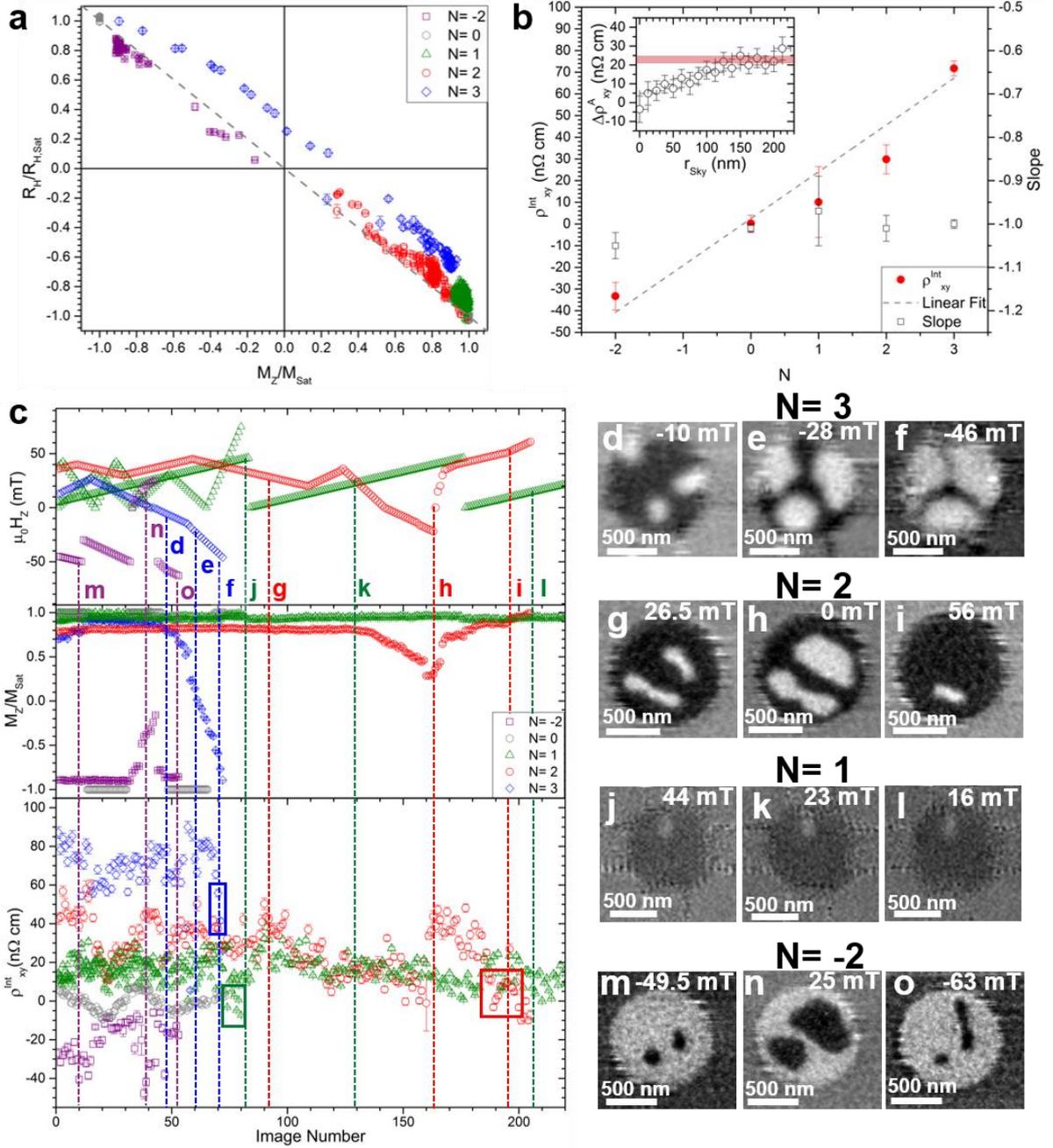

**Figure 3: Hall resistance signals from a single skyrmion. a**, Hall signal arising from the spin texture $R_H = R_{xy} - R_N$, normalised to the saturation value $R_{H,Sat}$, plotted here against the normalised magnetisation as determined from the STXM images, $M_Z/M_{Sat}$, for the case of saturation, $N = 0$, one skyrmion, $N = 1$, two skyrmions, $N = 2$, two skyrmions of opposite magnetisation, $N = -2$, and three skyrmions, $N = 3$. The grey dashed line shows the expected resistance change with magnetisation if there is only an anomalous Hall resistance contribution proportional to the magnetisation. **b** Intercept resistivity, $\rho^{Int}_{xy}$, and slope versus the number of skyrmions extracted from linear fits to the data shown in **a**. The best linear fit to $\rho^{Int}_{xy}$ versus $N$ reveals a contribution of 22±2 nΩ cm per skyrmion. The inset shows the anomalous Hall resistivity change from saturation, $\Delta\rho^A_{xy}$, versus measured skyrmion radius $r_{Sky}$ in the single skyrmion case. At radii below 150±30 nm, $\rho^{Int}_{xy}$ and $\rho^A_{xy}$, are of comparative scale. **c** Out of plane magnetic field, $M_Z/M_{Sat}$, and $\rho^{Int}_{xy}$, versus image number. The square boxes indicate magnetisation states were skyrmions states are annihilated. **d** to **o** are XMCD images at selected points, dashed lines, throughout the minor loops shown in **c**.

**METHODS**

The thin films were deposited by dc magnetron sputtering in a vacuum system with a base pressure of $2 \times 10^{-8}$ mbar and a target-substrate separation of roughly 7 cm. An argon gas pressure of 6.7 mbar was used during the sputtering and typical growth rates of around 0.1 nm/s were achieved. The superlattice stack, [Co (0.5 nm)/Ir (0.5 nm)/Pt (1.0 nm)]$_{\times 10}$, was grown on a seed layer of Ta (3.5 nm)/ Pt (3.8 nm) and capped with Pt (3.2 nm). The patterned structures were grown on 200 nm thick $Si_3N_4$ membranes supported on silicon (Silson Ltd, Warwickshire, UK). An identical superlattice stack was simultaneously sputtered on thermally oxidised silicon (with oxide layer thickness of 100 nm) to provide a sample for characterisation of material properties. X-ray reflectivity was used to measure the individual layer thicknesses on this sample. Room temperature polar magneto-optical Kerr effect magnetometry and in-plane superconducting quantum interference device vibrating sample magnetometry was used to confirm the out-of-plane easy axis of the superlattice (See supplementary information Fig. S1 (a)). The ordinary Hall coefficient, $R_0$, was measured to be -(1.9±0.2)×10$^{-11}$ Ωm/T using a 800 nm wide Hall bar (see supplementary information Fig. S1(b)).

1000 nm diameter discs were fabricated using multiple step electron beam lithography (see supplementary information Fig. S1(c) for scanning electron microscopy image of one of the devices measured). The lithographic recipe is described in more detail in the supplementary information.

The magnetic domains in the nanodiscs were imaged using scanning transmission x-ray microscopy (STXM) at the PolLux (X07DA) beamline at the Swiss Light Source using x-ray magnetic circular dichroism (XMCD) to provide magnetic contrast [47]. A spatial resolution of the order of 30 nm was achieved by using a Fresnel zone plate with an outermost zone of 25 nm to focus the x-rays on the sample. The images were taken at room temperature with the option of an out-of-plane quasi-static magnetic field. The sample surface was perpendicular to the incident x-rays, which were tuned to the Co $L_3$ absorption edge (ca. 778 eV). The x-ray magnetic dichroism contrast was measured by taking the difference between the absorption of left and right circularly polarized x-rays and dividing it by the sum of the absorption images. This leads to a black and white contrast indicating magnetic moments aligned parallel or antiparallel to the incident x-rays (see supplementary Fig. S1(d) for the field driven switching between the antiparallel saturation states and the corresponding Hall resistivity). The chamber was pumped down and backfilled with oxygen gas (5 mbar) to prevent carbon deposition. The disc was initially saturated at ±110 mT. Magnetic domains were nucleated using five bipolar 500 mV/-400mV 5 ns current pulses separated by 200 ns with an approximate current density of $7 \times 10^{11}$ A/m$^2$. The current pulses were created using a Tektronix AWG7122C arbitrary waveform generator. The skyrmion was then created by applying a field opposing the nucleated domain magnetisation alignment until only the most stable magnetic entity was left, the skyrmion (see supplementary information Fig. S2). Zero field stability of this skyrmion was observed. The electrical Hall and longitudinal resistance response to expanding and shrinking the skyrmion was observed by increasing and decreasing the field in a minor loop, with a STXM image acquired corresponding to each magnetotransport data point. The field dependent diameter evolution was extracted by fitting a circle to the XMCD contrast of the skyrmions which were observed to be radially symmetric; the error was given by the spatial resolution of the measurement.

The Hall resistance of the nanodisc was measured using a four probe approach where the Hall voltage and current electrodes overlapped at the edge of the nanodisc. The longitudinal resistance used a quasi-four probe method where the electrodes were offset by 1 μm from the disc (See supplementary Fig. S1 (c)). The electrodes were 500 nm in width and overlapped the edge of the disc by around 100 nm. A 17 Hz AC voltage was applied to the samples as well as a 250 mV DC voltage. The current was monitored using a 10 kΩ series resistor, R (see supplementary Fig. S1(c) inset). The series resistor voltage drop and Hall voltage was measured using a lock-in amplifier (Zurich

Instruments model HF2LI) (see Fig. 1 (c) inset for a circuit diagram). The magnetoresistance was measured using a Stanford Research Systems SIM910 in conjunction with a Keithley 2000 multimeter. The Hall resistance changes in the *N* = ±2 case were measured with a Keithley 2181A Nanovoltmeter. A Keithley 6221 DC and AC current source was used to apply 100 µA and the magnetoresistance was measured using a Keithley 2400. These measurements were performed with the sample mounted inside the PolLux STXM.

A Keithley 2182 Nanovoltmeter and a Keithley 6221 DC and AC current source was used to measure the Hall voltage and magnetoresistance of the 800 nm hall bar at 285 K. The field sample was placed in a solenoid superconducting magnet to enable the application of fields up to 8 T perpendicular to the sample.

# Discrete Hall resistivity contribution from Néel skyrmions in multilayer nanodiscs – supplementary information

Katharina Zeissler[*], Simone Finizio, Kowsar Shahbazi, Jamie Massey, Fatma Al Ma'Mari, David M. Bracher, Armin Kleibert, Mark C. Rosamond, Edmund H. Linfield, Thomas A. Moore, Jörg Raabe, Gavin Burnell, and Christopher H. Marrows

*Correspondence to k.zeissler@leeds.ac.uk

## Sample fabrication

1000 nm diameter discs were fabricated using multiple step electron beam lithography. All lithography was performed using a JEOL JBX 6300-FS at 100 kV accelerating voltage. First, small 60 µm squares were placed on the membrane (which would later be used to form the discs). In the same exposure alignment marks and large square regions were patterned onto the Si frame. A bilayer resist was used, comprising a bottom layer of copolymer methyl methacrylate (100 nm) and top layer of polymethyl methacrylate (200 nm) baked at 180°C for 5 min. A proximity effect correction algorithm was applied to the patterns on the Si frame and they were assigned a base dose of 440 µC cm$^{-2}$. The square patterns on the membrane were designed as "high" dose inner squares with 200 nm wide "low" dose outer contours (exposed at 530 and 175 µC cm$^{-2}$ respectively). This low-dose contour was necessary to force an undercut in the resist bilayer, since on the membrane, backscatter electrons are virtually non-existent. The patterns were written at a beam current of 2.0 nA and shot pitch of 6 nm. They were developed in a 3:7 ratio (by volume) of deionised water and isopropyl alcohol for 90 s, then rinsed in isopropyl alcohol for 30 s. The multilayer stack was then sputtered and lift off was performed in acetone followed by an isopropyl alcohol rinse.

The discs were then fabricated using a second electron beam lithography step, evaporated metal hard mask and ion milling. ZEP 520A (diluted 1:1 by volume in anisole) was spun to a thickness of 120 nm and baked at 180°C for 3 min. Using the alignment structures previously fabricated on the Si frame, the discs and Hall bar were exposed (343 µC cm$^{-2}$, shot pitch 5 nm, 1.0 nA beam current) and developed in n-amyl acetate for 60 s and rinsed in isopropyl alcohol for 30 s. New alignment structures, designed to overlay the large squares of multilayer on the Si frame (proximity effect corrected data, base dose 190 µC cm$^{-2}$), were also defined during this exposure. A metal hard mask was then thermally evaporated (Al (20 nm)/ Ti (15 nm)) and lifted off in cyclopentanone at 65°C. The Ti has a low sputter yield so is suitable for masking during the mill whilst the Al can be easily dissolved to remove the mask after patterning. A 2 min argon ion mill was used to shape the nanostructures and new alignment marks. The milling was performed at a beam current of 30 mA, beam voltage of 1000 V, an acceleration voltage of 100 V and an argon flow rate of 5 sccm which gave a working pressure of around 4×10$^{-5}$ mbar (the base pressure of the system was in the low 10$^{-7}$ mbar). The sample was milled at a 30° angle which was increased to 55° for the last 15 s. After milling, the mask remainder was removed in MF319 developer which dissolves the Al layer.

The Ti (5 nm)/ Au (65 nm) electrodes, in case of the two discs resulting in the $N$ = 1 and $N$ = 3 measurements, and the Ti (5 nm/ Cu (65 nm) electrodes, in case of the $N$ = ±2 disc, were formed by a third electron beam lithography step, thermally evaporation and lift off. A bi-layer resist comprising copolymer methyl methacrylate (200 nm) and polymethyl methacrylate (200 nm) were spun and baked at 180°C for 5 min. To reduce the write time, the electrode design was divided into inner (fine) and outer (coarse) contacts. The inner contacts (exclusively on the membrane) were written at 500 pA, 4 nm shot pitch and 660 µC cm$^{-2}$. The outer contacts (which crossed from the membrane to the frame) were exposed using a 10.0 nA beam at 20 nm shot pitch. They consisted of a 2 µm wide high dose outer "sleeve" and proximity corrected inner region, the former was exposed at

660 µC cm$^{-2}$ whilst the latter had a base dose of 440 µC cm$^{-2}$. The pattern was developed in a 3:7 ratio by volume of deionised water and isopropyl alcohol for 150 s and then rinsed with isopropyl alcohol for 30 s. The interface of the disc yielding the $N$ = ±2 data was cleaned of resist residue with a 4 min ultraviolet-ozone treatment prior to the electrode evaporation. After evaporation, the structures were lifted off in acetone and rinsed with isopropyl alcohol.

**Sample characterisation**

Three discs of 1 µm diameter were fabricated from [Co (0.5 nm)/Ir (0.5 nm)/Pt (1.0 nm)]$_{×10}$ multilayers using electron beam lithography and Ar ion milling, along with a Hall bar with a width of 800 nm. Fig. S1a shows the out of plane magneto-optical Kerr and in-plane superconducting quantum interference device vibrating sample magnetometry measurements on simultaneously deposited sheet films in Fig. S1a confirming out-of-plane magnetic anisotropy. The Hall bar was used to measure the ordinary Hall coefficient, $R_0$, of the superlattice stack (see Fig. S1b for the Hall and magnetoresistivity $\rho_{xy}$ and $\rho_{xx}$ respectively); $R_0$ was measured to be -(1.9±0.2)×10$^{-11}$ Ωm/T by a linear fit to $\rho_{xy}$ for fields exceeding 1 T. The longitudinal resistivity was measured to be virtually field independent; no more than a 0.005% change was observed over 8 T. A scanning electron microscopy image of one of the discs is shown in Fig. S1c. The longitudinal voltage, Hall voltage and current electrodes attached to the nanodisc are identified by $V_{xx}$, $V_{xy}$, and $I$ respectively. A schematic of the electrical measurement set-up is shown in Fig. S1c inset. Care was taken to accurately determine and subtract the small offset observed in the Hall resistivity due to a small longitudinal contribution caused by non-perfect electrodes (0.6 % of the measured $\rho_{xx}$). This small offset was determined by comparing the resistivity values at positive and negative magnetic saturation after each skyrmion measurement was concluded.

Fig. S1d shows these Hall resistivity measurements at different current densities while an out-of-plane field was stepped between ±110 mT. A sharp reversal between well-defined saturated Hall resistivity levels of ±220±10 nΩ cm was observed. The disc was seen to switch from the positive saturated state (black XMCD contrast) to the negative saturated state (white XMCD contrast) within one field step by STXM (see Fig. S1d), confirming that this Hall signal is that from a uniformly magnetised disc. The coercive field was found to be around 89 mT. The domain wall speed of Pt/Co/Ir trilayers, at such high fields, was observed to be of the order of 50 µm/s$^4$. Hence, the nucleated domains travel through the 1 µm disc much faster than the magnetic state was imaged (acquisition time is of order of a few minutes per helicity). The inset in Fig. S1d shows a linear relationship between the applied AC voltage and AC current passed through the disc.

**Single skyrmion nucleation**

Magnetisation textures were created within the disc by following a magnetic field and current pulse nucleation protocol. The disc was initially saturated at +110 mT, the corresponding STXM image is shown in Fig. S2a. The field was then reduced back to 0 mT and five consecutive bipolar current pulses were applied to the nanodisc. Both the negative and the positive part of the bipolar pulse had a duration of 5 ns. The pulse amplitude was 500 mV followed by a -400 mV part. Each of the 10 ns bipolar pulses were separated by 200 ns. The positive current pulse direction is indicated by the red arrow in Fig. S2a. This corresponds to a current density of roughly 7×10$^{11}$ A/m$^2$. The current was seen to induce a complex magnetic domain state inside the nanodisc, most probably through Joule heating (Fig. S2b). The resulting magnetic state could then be altered by applying a positive out-of-plane field. The field applied opposed the magnetisation of the current nucleated domains which lead to their shrinking. As the domain state of the disc is altered (see Fig. S2c to f) the changes in the Hall resistivity were measured (see Fig. S2g). The changes in the Hall resistivity are accompanied by changes in normalised out-of-plane magnetisation $M_z/M_{sat}$ (see Fig. S2h). The normalised magnetisation was computed from the XMCD contrast images by counting the number of black, white and grey pixels in the disc and assigning them a value of 1, -1 and 0 and normalising the sum by the number of pixels in the disc. The average grey value of the non-magnetic membrane surrounding the disc was used as the counting threshold (see details in the supplementary information). Just before saturation, at around 43 mT, a single skyrmion (an $N$ = 1 state) can be observed in the STXM images (Fig. S2e and f). The skyrmion at 46 mT in Fig. S2f is shown again in Fig. 1d of the main manuscript after the field was reduced to 0 mT.

**Magnetic contrast image analysis**

The XMCD-STXM images which were used to extract the normalised magnetisation of the Pt/Co/Ir discs at each step of the magnetic field reversal were acquired using an Au Fresnel zone plate to focus the x-rays. The outermost zone width of the zone plate was 25 nm. The entrance and exit slits to the monochromator of the PolLux beamline were tuned to achieve an x-ray spot diameter of approximately 50 nm. This was found to be the best compromise of imaging resolution while retaining a high photon flux. The secondary source is thus not point-like and the point spread function of the Fresnel zone plate can be described with a Gaussian function exhibiting a variance of approximately 25 nm [1].

To extract the magnetization state the XMCD-STXM images, acquired using a point resolution between 15 nm (for the $N$ = ±2 states) and 25 nm (for the $N$ = 1 and $N$ = 3 states), which represent a convolution of the point spread function of the Fresnel zone plate and the magnetization state, were deconvoluted using an iterated Lucy-Richardson algorithm [1]. Prior to the deconvolution the images were padded at the edges to reduce artefacts. The deconvolution of the XMCD-STXM images leads to an improvement of the spatial resolution and contrast, which is particularly evident when observing the domain walls (see Fig. S3 a - c).

The normalized magnetization was calculated from the deconvoluted XMCD-STXM images through a thresholding algorithm (see Fig. S3 d - g for an example). A positive pixel value (white in Fig. S3d) in the XMCD-STXM indicates a negative $M_z$ (i.e. pointing out of the page), a negative value (black in Fig. S3d) indicates a negative $M_z$ (i.e. pointing into the page), and 0 indicates either non-magnetic or in plane magnetized regions. Therefore, to determine the normalized magnetization $M_z/M_{Sat}$, the pixels of the different areas were added assigning a value of +1, -1 and 0 to black, white and grey pixels respectively and normalized it to the number of pixels in the disc. The area of the disc was determined from elemental images of the discs, where only the regions exhibiting elemental contrast at the Co $L_3$ edge were considered. The gray threshold was determined by considering the variance of the noise in the non-magnetic areas outside of the Pt/Co/Ir disc; resulting in a threshold boundary of ±0.08.

The $M_z/M_{Sat}$ value was extracted using Matlab and Python scripts. Pre-analysis the images of each dataset were registered to the first image in the dataset. The registration was carried out on the topographic images (obtained by calculating the sum of the positive and negative helicity images). This prevents registration errors due to the changes in the magnetic configuration of the Pt/Co/Ir discs. The registration of the images was carried out first by determining the edges of the discs by applying a Sobel edge filter to the images and then by applying an optimized cross-correlation algorithm with sub-pixel accuracy [2].

The counting error was evaluated using counting statistics, i.e. counting an area of $M_z = +1$ as $M_z = -1$ and vice versa. If we consider counting black pixels the error in $M_z$ is given by $\sqrt{n}/n_{Total}$ where $n$ is the number pixels counted as $M_z=+1$ and $n_{Total}$ is the number of all pixels in the disc.

**Imaging of the Néel domain walls**

Fig. S4 shows x-ray photoemission electron microscopy XMCD images taken at the Surface Interface Microscopy beamline of the Swiss Light Source [3]. Imaging was conducted at the Co $L_3$ edge on a continuous film of Pt/Co/Ir multilayer with an identical stack to that which was studied by STXM, grown on a Si substrate in the same growth run. The 16 degree (with respect to the sample plane) X-ray incidence angle provides sensitivity to both the out-of-plane magnetization and the in-plane magnetization from the XMCD contrast. The images were separated into the pure out-of-plane and in-plane contrast by acquiring two XMCD- x-ray photoemission electron microscopy images with the sample rotated by 180 degrees about its normal between them (see Fig. S4a and b respectively). After the addition or subtraction of the two images, one obtains out-of-plane or in-plane magnetic contrast, respectively (see Ref. [4] for details on the determination of the domain wall type from the XMCD- x-ray photoemission electron microscopy images). The line scan taken at the same point through both XMCD contrast images shows that the domain walls are of the left-handed Néel type (see Fig. S4c).

**Current flow distortions caused by skyrmions**

The redistribution of the current due to the presence of a skyrmion was simulated using the *Finite Element Method Magnetics* (FEMM 4.2) open source simulation package developed by Aladdin Enterprises [5]. The influence of the skyrmion on the current distribution was evaluated by creating regions of changed conductivity due to the in-plane Néel domain wall separating the skyrmion from the surrounding area. The Néel domain wall, which forms the boundary of the skyrmion, results in an in-plane magnetization component and rotates locally from being parallel to the current flow to being perpendicular to the current flow at the front/back and sides of the skyrmion, respectively. Hence, there are two regions of higher resistance on either end of the skyrmion due to the anisotropic magnetoresistance effect. The anisotropic magnetoresistance measured on a similar 800 nm wide Hall bar on silicon Ta (4.4 nm)/ Pt (3.1 nm)/[Co (0.6 nm)/Ir (0.4 nm)/Pt (1.1 nm)]$_{\times 10}$/Ta (2.2 nm) reveals a maximum anisotropic magnetoresistance ratio of 0.04 % (See Fig. S5a). Fig. S5b shows the simulated planar Hall resistivity without a skyrmion, for a skyrmion of 100 nm diameter placed at the disc centre (Fig. S5c) and placed at the edge (Fig. S5 d) (roughly in the same position as the skyrmion imaged in Fig. 1d of the main manuscript). The multilayer system was simplified to a single 5 nm cobalt layer with a conductivity (when saturated) of $1.7 \times 10^7$ S/m and a change of 0.05 %, 1 %, 10 %, 20 % and 25 % in the skyrmion region. In order to measure a planar Hall resistivity of 18 nΩ cm, the anisotropic magnetoresistance would have to be 15 %; this is much larger than we measure and a highly unrealistic value for a transition metal ferromagnet.

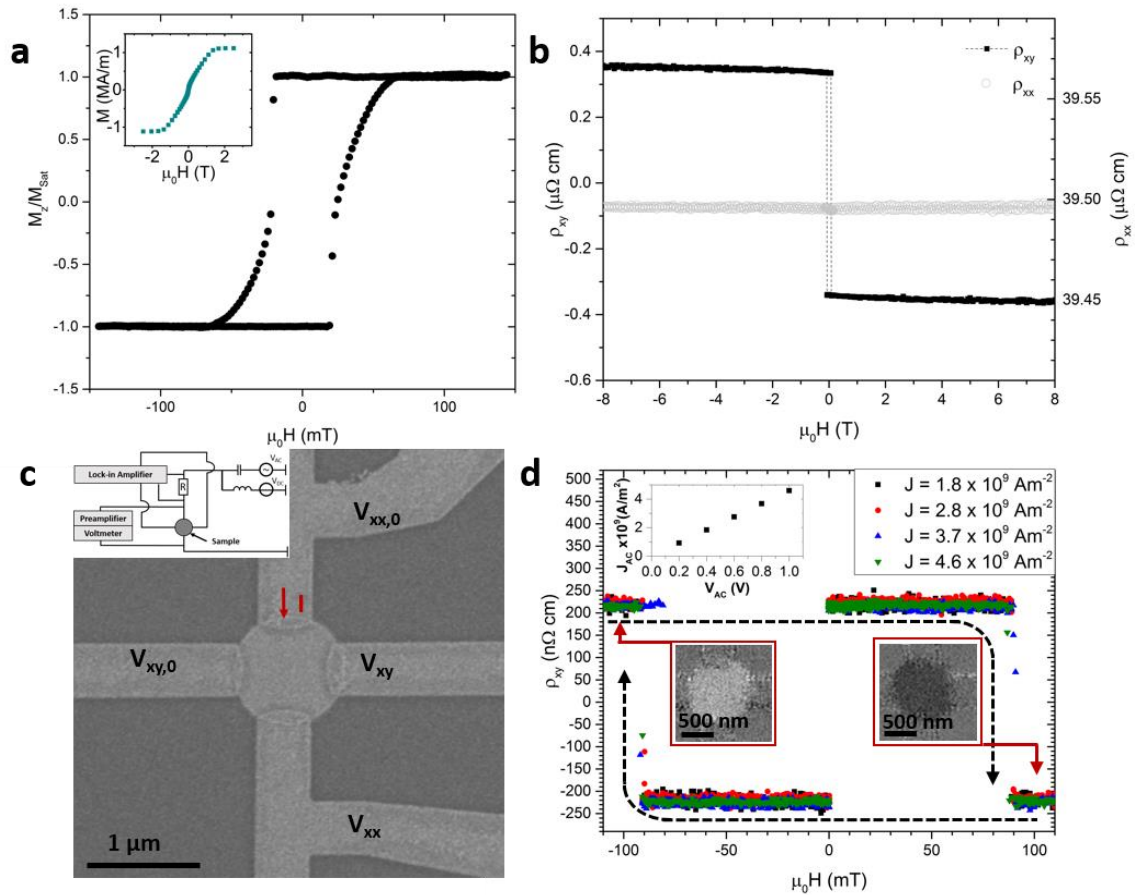

**Figure S1: Sheet-film magnetic properties of [Pt/Co/Ir]$_{\times 10}$ multilayer and experimental set up. a**, Out-of-plane magnetisation versus magnetic field loop acquired via Kerr magnetometry of an unpatterned film. Inset shows SQUID vibrating sample magnetometry measurements of magnetisation versus in plane magnetic field. The easy axis of the superlattice structure is out-of-plane. **b**, Electrical transport measurement of Hall and longitudinal resistivities of the 800 nm wide Hall bar as a function of out-of-plane magnetic field. **c**, Scanning electron microscopy image of one of the fabricated nanodiscs. The current, magnetoresistance, and Hall resistance, electrodes are annotated with $I$, $V_{xx}$ and $V_{xx,0}$, and $V_{xy}$ and $V_{xy,0}$, respectively. The inset shows the electrical measurement circuit diagram. **d**, Hall resistivity, of a nanodisc as a function of field on a major hysteresis loop at different current densities $J$. The black dashed arrows show the field sweep direction. The inset shows the linear relationship between $J$ and the measured AC voltage $V_{AC}$. No magnetic domains could be observed within the disc in this major loop, as shown in the inset STXM images.

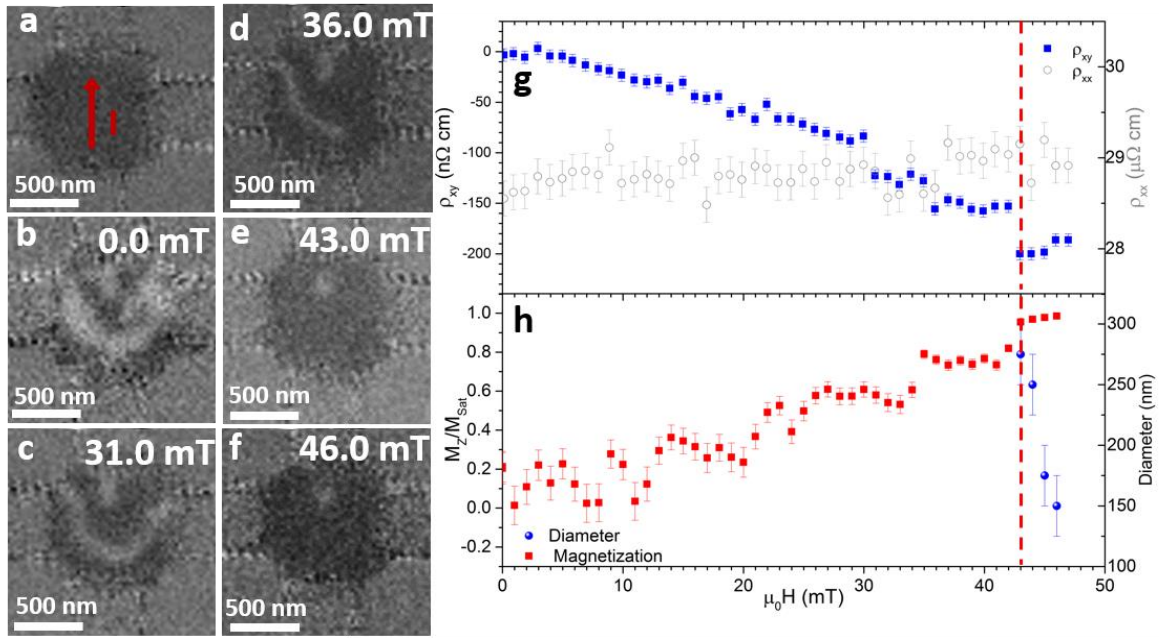

**Figure S2: Skyrmion nucleation by field and current pulses. a**, After initial saturation at +110 mT, a train of current pulses was transmitted in the direction of the red arrow at 0 mT, **b**, leading to the formation of a complex domain pattern. Dark and light regions of XMCD contrast show antiparallel magnetised domains. **c-f**, After an out-of-plane field was applied the nucleated domains were observed to decrease in size. The applied field was aligned antiparallel to the magnetisation direction at the core of the nucleated domains. At high fields, **e** and **f**, the most stable magnetic object remains, a skyrmion. **g**, Hall resistivity, $\rho_{xy}$, magnetoresistivity, $\rho_{xx}$, and **h**, magnetisation changes as the initial domain state was altered using an out-of-plane magnetic field. The magnetisation was extracted from the ratio of dark and bright regions of XMCD contrast in the STXM images. The skyrmion diameter was determined from fitting a circle to the image. The red dashed line shows the field above which the last domain collapses into the single skyrmion.

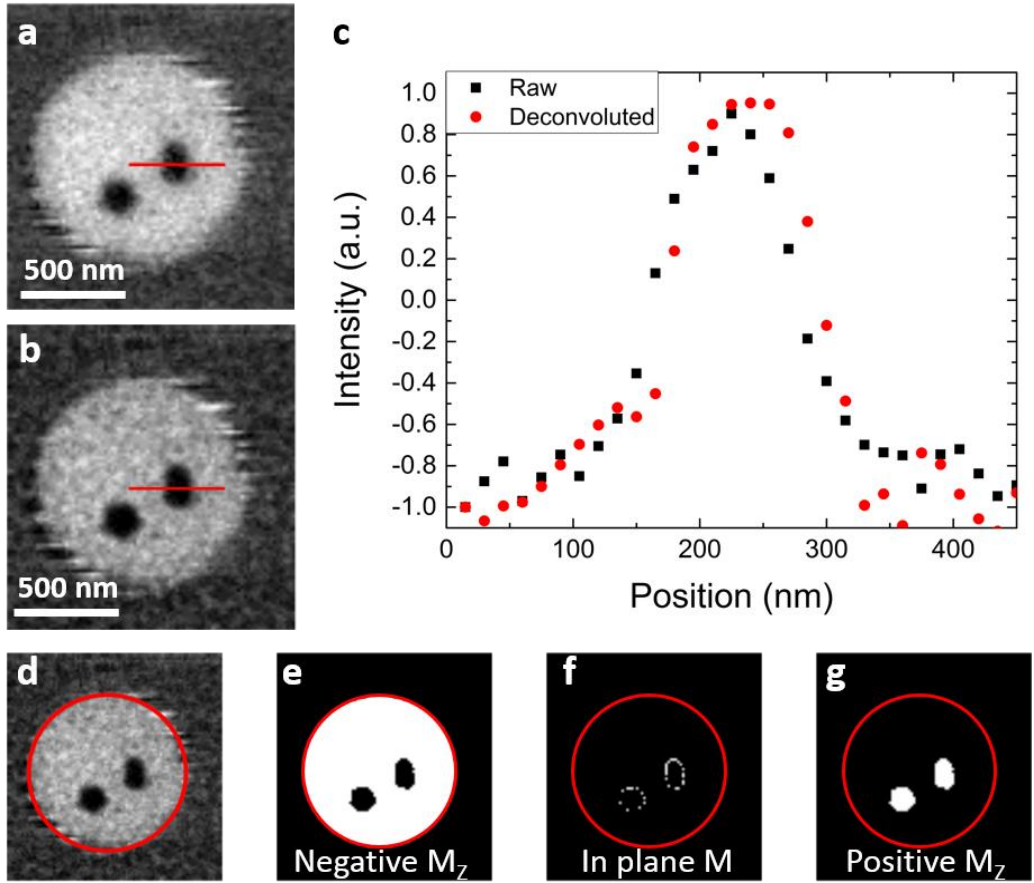

**Figure S3: Normalised magnetisation extraction from the magnetic contrast images.** (a) Raw and (b) deconvoluted XMCD-STXM image of a 1 µm wide Pt/Co/Ir disc exhibiting two magnetic skyrmions. (c) Corresponding line scan taken across one magnetic skyrmions (marked by the red line in (a) and (b)) The domain wall width of the raw images was found to be approximately 75 nm, in comparison the domain wall width of the deconvoluted image was found to be approximately 45 nm. (d) XMCD-STXM shown in (b) in which the red circle encloses the area considered for the pixel counting. Analysed image showing the pixels counted as (e) "$M_z = -1$", (f) "$M_z = 0$" and "$M_z = +1$" in white.

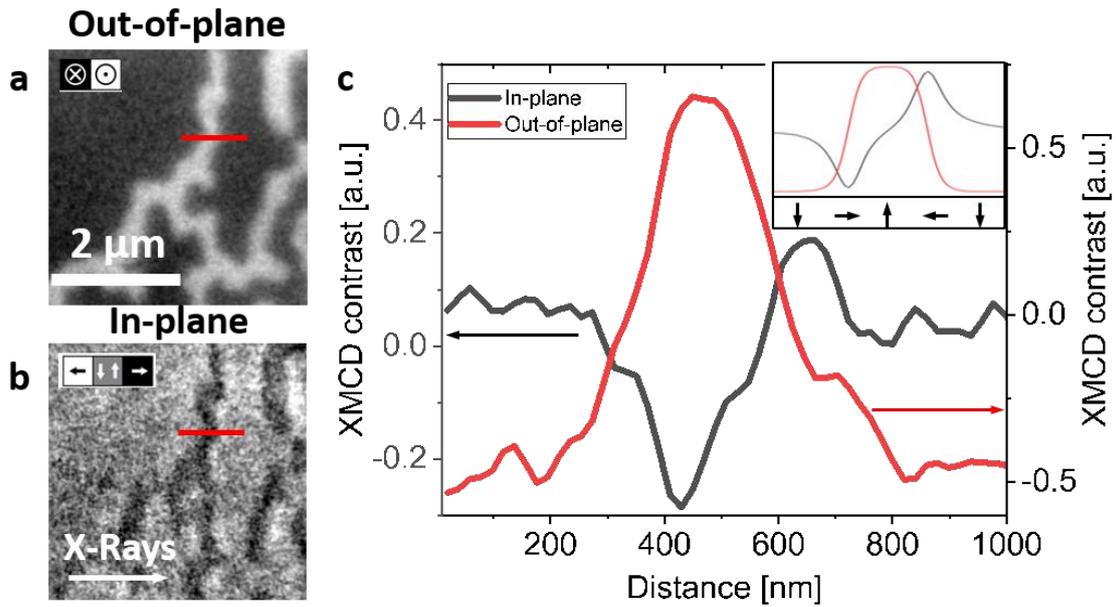

**Figure S4: In plane moment sensitive magnetic images**. XMCD contrast imaged using x-ray photoemission electron microscopy of **a** the out-of-plane and **b** the in-plane magnetic configuration in a continuous film of a multilayered Pt/Co/Ir sample. The direction of the probing x-ray beam is along the horizontal axis and the grayscale bars in **a** and **b** indicate the direction of the magnetic contrast in the images. **c** The line profile is consistent with a left-handed Néel domain wall; Inset shows a sketch of the expected out of plane and in plane line profile for a left-handed Néel domain wall.

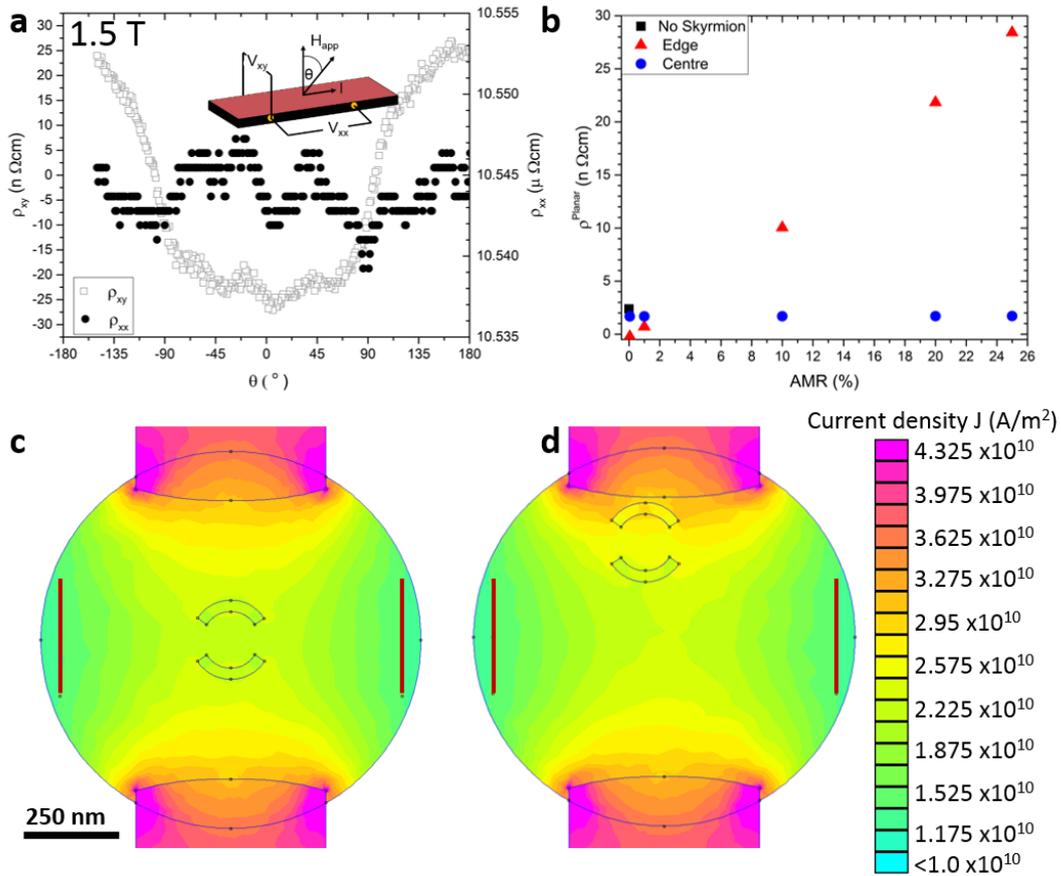

**Figure S5: Planar Hall effect contribution to the Hall resistivity**. **a** Angular dependent resistivity measured on an equivalent 800 nm Hall bar. A maximum anisotropic magnetoresistance of 0.04 % was found. **b** Planar Hall resistivity simulated using FEMM for a disc of 7 nm cobalt without the presence of a skyrmion, with a 100 diameter skyrmion at its center and with a skyrmion at the edge and away from the mirror axis of the disc for different anisotropic magnetoresistance ratios. **c** and **d** shows the simulated current density throughout the disc when a skyrmion is present at the center and at the edge respectively for an unrealistically large anisotropic magnetoresistance ratio of 20 %. The red lines show the breadth over which the voltage was averaged to calculate the planar Hall resistivity shown in **b**.